\documentclass[12pt]{article}

\usepackage{hyperref}
\usepackage{cite}
\usepackage{units}
\usepackage{epsfig}
\usepackage{pifont}
\usepackage{amsmath,amsfonts,amssymb,amsthm,mathrsfs}
\usepackage{graphicx,chngcntr,slashed}
\usepackage{cancel}
\usepackage{pstricks}
\usepackage{afterpage}
\usepackage{braket}
\usepackage{caption}
\usepackage{subcaption}
\usepackage{xcolor}
\usepackage{verbatim}
\usepackage{multirow}
\usepackage{tabularx,caption}
\usepackage{float}
\usepackage{afterpage}
\usepackage{comment}

\def\mathswitch#1{\relax\ifmmode#1\else$#1$\fi}

\newcommand{\mht}{\hat{t}}
\newcommand{\mhu}{\hat{u}}
\newcommand{\mhs}{\hat{s}}

\newcommand{\mycaption}[1]{\caption{\sl #1}}

\makeatletter
\def\section{\@startsection {section}{1}{\z@}{+3.0ex plus +1ex minus
  +.2ex}{2.3ex plus .2ex}{\large\bf\boldmath}}
\def\subsection{\@startsection{subsection}{2}{\z@}{+2.5ex plus +1ex
minus +.2ex}{1.5ex plus .2ex}{\normalsize\bf\boldmath}}
\def\subsubsection{\@startsection{subsubsection}{3}{\z@}{+3.25ex plus
 +1ex minus +.2ex}{1.5ex plus .2ex}{\normalsize\it}}

\graphicspath{{.}{plots/}}

\begin{document}
\thispagestyle{empty}

\def\thefootnote{\fnsymbol{footnote}}

\begin{flushright}
\end{flushright}

\vspace{1cm}

\begin{center}

{\Large {\bf Disentangling SMEFT operators with future low-energy PVES experiments}}
\\[3.5em]
{\large
Radja~Boughezal$^1$, Frank~Petriello$^{1,2}$ and Daniel~Wiegand$^{1,2}$
}

\vspace*{1cm}

{\sl
$^1$ HEP Division, Argonne National Laboratory, Argonne, Illinois 60439, USA \\[1ex]
$^2$ Department of Physics \& Astronomy, Northwestern University,\\ Evanston, Illinois 60208, USA
}

\end{center}

\vspace*{2.5cm}

\begin{abstract}
We study the potential of future Parity-Violating Electron Scattering (PVES) data to probe the parameter space of the Standard Model Effective Field Theory (SMEFT). We contrast the constraints derived from Drell-Yan data taken at the Large Hadron Collider (LHC) with projections of the planned PVES experiments SoLID and P2. We show that the PVES data can complement the bounds set by the LHC data in the dimension-6 operator space since it probes different combinations of operators than Drell-Yan. The lower characteristic energy of P2 and SoLID also helps disentangle effects of dimension-6 and dimension-8 operators that are difficult to resolve with LHC Drell-Yan data alone.
\end{abstract}

\setcounter{page}{0}
\setcounter{footnote}{0}

\newpage


\section{Introduction}

No conclusive sign of physics beyond the Standard Model (SM) has been observed at either the Large Hadron Collider (LHC) or in other experiments. This has motivated significant attention toward studying how potential extensions of the SM are constrained by the current data. Understanding the implications of current measurements will provide insight on both the energy scale at which new physics can appear and to what sectors of the SM new heavy particles are allowed to couple given the existing constraints.

A consistent framework to perform such an analysis, with the assumption that any new physics is significantly heavier than the electroweak scale, is the SM Effective Field Theory (SMEFT). The SMEFT is constructed by augmenting the SM Lagrangian with higher-dimensional operators consistent with the SM gauge symmetries and formed only from SM fields. The higher-dimensional operators in the SMEFT are suppressed by appropriate powers of a characteristic energy scale $\Lambda$ below which heavy new fields are integrated out. Complete, non-redundant bases for the dimension-6~\cite{Buchmuller:1985jz,Arzt:1994gp,Grzadkowski:2010es} and dimension-8 operators~\cite{Murphy:2020rsh,Li:2020gnx} have been identified. Odd-dimensional operators violate lepton-number and will not be considered in this work.

It is an ongoing effort to analyze the numerous available data within the SMEFT framework, primarily in partial analysis of individual SMEFT sectors~\cite{Han:2004az,Chen:2013kfa,Ellis:2014dva,Wells:2014pga,Falkowski:2014tna,Cirigliano:2016nyn,deBlas:2016ojx,Hartmann:2016pil,Falkowski:2017pss,Biekotter:2018rhp,Grojean:2018dqj}. Recent work has been devoted to performing a global, simultaneous fit of all data available~\cite{Pomarol:2013zra,DiVita:2017eyz,Almeida:2018cld,Ellis:2018gqa,Hartland:2019bjb,Brivio:2019ius,vanBeek:2019evb,Aoude:2020dwv,Ellis:2020unq}. Most of these global fits have focused on the truncation of the SMEFT expansion to dimension-6 operators. Several open issues must be confronted when performing such analyses. When performing a global fit with a greater numbers of parameters than observables, only certain linear combinations of the fit parameters can be probed. These flat directions in parameter space can be either exact or approximate in the sense that they hold only in certain kinematic limits. Large deviations from the SM, as measured by turning on SMEFT operators with sizeable Wilson coefficients, can lead to no observable consequences in the presence of such degeneracies. Finding observables that remove these flat directions is an important component of improving the efficacy of global fits to the SMEFT. Another issue to address is the sensitivity of fits to dimension-8 and higher operators. Intuitively their effects should be suppressed, but since many measurements at the LHC probe high energies this assumption must be tested. Furthermore, dimension-8 effects sometimes represent the leading SMEFT contributions in models with certain approximate symmetries~\cite{Liu:2016idz,Contino:2016jqw}. Previous analyses of the impact of dimension-8 operators can be found in the literature~\cite{Degrande:2013kka,Hays:2018zze,Bellazzini:2018paj,Ellis:2019zex,Alioli:2020kez,Murphy:2020cly,Hays:2020scx,Ellis:2020ljj}.

The goal of this paper is to investigate the potential impact of high-luminosity, low-energy parity violating electron scattering (PVES) experiments in resolving both issues discussed above. Our work is motivated by the planned SoLID experiment at Jefferson Laboratory~\cite{Wang:2014bba,Wang:2014guo} as well as the planned P2 experiment at MESA~\cite{Becker:2018ggl}. These two low energy experiments are complementary to each other since, due to the nature of the targets used, they probe different combinations of Wilson coefficients. Our analysis focuses on semi-leptonic four-fermion operators at both dimension-6 and dimension-8. Previous analyses~\cite{Falkowski:2017pss} of PVES data in terms of the asymmetry parameter $A_{PV}$ focused on constraining dimension-6 modifications of the electroweak couplings. Certain combinations of dimension-6 semi-leptonic four-fermion operators are already well-probed by high invariant mass Drell-Yan distributions at the LHC~\cite{Falkowski:2017pss,Dawson:2018dxp,Alte:2018xgc}. In principle the available LHC data should also be able to constrain the corresponding dimension-8 operators due to the large integrated luminosity that has been collected as well as the sufficiently large center of mass energy. In practice the Drell-Yan process exhibits numerous flat directions that complicates the separation of different dimension-6 effects~\cite{Alte:2018xgc,Boughezal:2020uwq}, as well the disentanglement of dimension-6 from dimension-8 operators as we show later in this manuscript. The considerably lower energy of the PVES experiments leads to a suppression of dimension-8 effects, and therefore sensitivity to dimension-6 operators only. Combining LHC with SoLID and P2 respectively allows these different order operators to be disentangled. We furthermore show that PVES experiments can be used to lift flat directions in the space of dimension-6 operators when combined with Drell-Yan data. Our work follows in the spirit of previous analyses that showed how future data from an electron-ion collider (EIC) could help resolve degeneracies present in SMEFT fits using Drell-Yan data only~\cite{Boughezal:2020uwq}. One advantage of the SoLID and P2 experiments is that they are anticipated to start data-taking within the next few years, as opposed to the longer time frame of the EIC.

Our paper is organized as follows. In Section~\ref{sec:smeft} we review the aspects of the SMEFT framework relevant for our analysis. We present and discuss the formulae describing the Drell-Yan process at the LHC and parity-violating scattering at SoLID and P2 in Section~\ref{sec:DYsolid}. In Section~\ref{sec:Comparison} we present the main results of our paper, combined fits of the Drell-Yan data with SoLID and P2 projections, and illustrate their potential to differentiate between both dimension-6 and dimension-8 operators. Finally, in Section~\ref{sec:conc} we put our findings in perspective and conclude.

\section{Notation and SMEFT formalism} \label{sec:smeft}

We review in this section aspects of the SMEFT relevant for
our analysis of LHC and projected PVES data. The SMEFT is an extension of the SM Lagrangian including terms
suppressed by an energy scale $\Lambda$ at which the ultraviolet completion is assumed to
become important and new particles beyond the SM appear. Truncating the expansion in $1/\Lambda$ at dimension-8, and
ignoring operators of odd-dimension which violate lepton number, we have
\begin{equation}
{\cal L} = {\cal L}_{SM}+ \frac{1}{\Lambda^2}\sum_i C^{6}_{i} {\cal O}_{6,i} + \frac{1}{\Lambda^4}\sum_i C^{8}_{i} {\cal O}_{8,i} + \ldots,
\end{equation}
where the ellipsis denotes operators of higher dimensions. The Wilson coefficients $C^{d}_{i}$ defined above are dimensionless. We calculate cross sections to leading order in the coupling constants as well as to dimension-8 in the SMEFT expansion. This includes contributions from both true dimension-8 operators as well as contributions of dimension-6 operators squared. For both SoLID and P2 observables we have explicitly checked that dimension-8 contributions are suppressed like $Q^2/\Lambda^4$, where $Q^2 < 6\,\textrm{GeV}^2$ is the energy transfer relevant for the SoLID and P2 experiments. Since the SMEFT requires $\Lambda$ to be much greater than the electroweak scale all dimension-8 effects are completely negligible for PVES kinematics. For notational simplicity we will drop the explicit dimension labels on the operators and Wilson coefficients when no confusion between them can occur.

In Table~\ref{tab:operators} we compile the operators that affect the Drell-Yan and PVES processes at leading order in the coupling constants. We have used the notation of Ref.~\cite{Murphy:2020rsh} for the dimension-8 operators. We assume massless fermions as well as minimal flavor violation for the SMEFT Wilson coefficients, and have neglected the scalar and tensor operators that vanish with these assumptions. We note at this point that for each dimension-6 operator there are two dimension-8 extensions that differ in the placement of the covariant derivatives. We will refer to those compiled in Table~\ref{tab:operators} as``type-1" operators. The other possibility, which we will call "type-2" can schematically be written as $(\overline{\psi} \gamma^\mu \overleftrightarrow{D^\nu}\psi)(\overline{\psi} \gamma_\mu \overleftrightarrow{D_\nu}\psi)$ in terms of the left-right derivative $\overleftrightarrow{D^\mu} = \overrightarrow{D^\mu} - \overleftarrow{D^\mu}$. As we will explain in greater detail later the type-1 operators lead to only an energy-dependent shift of the corresponding dimension-6 effects, while the type-2 operators lead to a different angular dependence. The effect of the type-2 dimension-8 extensions can therefore in principle be disentangled through angular variables~\cite{Alioli:2020kez} and we omit them from our analyses.
\begin{table}[h!]
\begin{center}
  \begin{tabular}{| c | c ||  c | c |}
  \hline
  \multicolumn{2}{|c||}{Dimension 6} & \multicolumn{2}{c|}{Dimension 8}\\
    \hline
    $\mathcal{O}_{lq}^{(1)}$ & $\left(\overline{l} \gamma^\mu l \right)\left(\overline{q} \gamma_\mu q \right)$ & $\mathcal{O}^{(1)}_{l^2q^2D^2}$ & $D^\nu \left(\overline{l} \gamma^\mu l \right) D_\nu \left(\overline{q} \gamma_\mu q \right)$ \\

    $\mathcal{O}_{lq}^{(3)}$ & $\left(\overline{l} \gamma^\mu\tau^i l \right) \left(\overline{q} \gamma_\mu\tau^i q \right)$ & $\mathcal{O}^{(3)}_{l^2q^2D^2}$  &$D^\nu \left(\overline{l} \gamma^\mu\tau^i l \right) D_\nu \left(\overline{q} \gamma_\mu\tau^i q \right)$\\ 

   $\mathcal{O}_{eu}$ &  $\left(\overline{e} \gamma^\mu e \right) \left(\overline{u} \gamma_\mu u \right)$  & $\mathcal{O}^{(1)}_{e^2u^2D^2}$  & $D^\nu \left(\overline{e} \gamma^\mu e \right) D_\nu \left(\overline{u} \gamma_\mu u \right)$\\ 

   $\mathcal{O}_{ed}$ & $\left(\overline{e} \gamma^\mu e \right) \left(\overline{d} \gamma_\mu d \right)$  & $\mathcal{O}^{(1)}_{e^2d^2D^2}$  & $D^\nu \left(\overline{e} \gamma^\mu e \right) D_\nu \left(\overline{d} \gamma_\mu d \right)$\\

  $\mathcal{O}_{lu}$ & $\left(\overline{l} \gamma^\mu l \right)\left(\overline{u} \gamma_\mu u \right)$  & $\mathcal{O}^{(1)}_{l^2u^2D^2}$ & $D^\nu \left(\overline{l} \gamma^\mu l \right) D_\nu \left(\overline{u} \gamma_\mu u \right)$ \\ 

    $\mathcal{O}_{ld}$ & $\left(\overline{l} \gamma^\mu l \right)\left(\overline{d} \gamma_\mu d \right)$  & $\mathcal{O}^{(1)}_{l^2d^2D^2}$ &  $D^\nu \left(\overline{l} \gamma^\mu l \right) D_\nu \left(\overline{d} \gamma_\mu d \right)$ \\

   $\mathcal{O}_{qe}$ & $\left(\overline{q} \gamma^\mu q \right) \left(\overline{e} \gamma_\mu e \right)$  & $\mathcal{O}^{(1)}_{q^2e^2D^2}$ & $D^\nu \left(\overline{q} \gamma^\mu q \right) D_\nu \left(\overline{e} \gamma_\mu e \right)$\\
   \hline
  \end{tabular}
    \mycaption{Dimension-6 operators that potentially contribute to the Drell-Yan process and their relevant dimension-8 extensions. We refer to the dimension-8 operators in this table as ``type-1". The corresponding ``type-2" operators are not shown.\label{tab:operators}}
\end{center}
\end{table}
In Table~\ref{tab:operators} $l$ and $q$ denote $SU(2)$ lepton and quark doublets respectively, while $e$, $u$ and $d$ are the right-handed singlets. $\tau^i$ are the $SU(2)$ Pauli matrices and $D_\mu$ is the covariant derivative. We suppress flavour indices for notational clarity. Though our analysis of PVES data is only sensitive to first-generation operators it is necessary to disentangle contributions from different generations when preparing a global fit of all operators. A possible strategy in the leptonic sectors was presented in Ref.~\cite{Boughezal:2020klp}. The Wilson coefficients are in principle dependent on the renormalization scheme chosen. In an $\overline{\text{MS}}$ scheme they become scale-dependent and run with energy. As we perform only a leading-order analysis in this work we neglect this running.

Historically it has been customary to parameterize the parity-violating, dimension-6 interactions in terms of the following phenomenological four-fermion Lagrangian~\cite{Zyla:2020zbs}:
\begin{align}
\mathcal{L}_{PV} = \frac{G_F}{\sqrt{2}}\bigg[&(\overline{e}\gamma^\mu \gamma_5 e) (C^6_{1u} \overline{u}\gamma_\mu u + C^6_{1d} \overline{d}\gamma_\mu d)+ (\overline{e}\gamma^\mu e) (C^6_{2u} \overline{u}\gamma_\mu\gamma_5 u + C^6_{2d} \overline{d}\gamma_\mu\gamma_5 d)  \nonumber\\
&+(\overline{e}\gamma^\mu e) (C^6_{Vu} \overline{u}\gamma_\mu u + C^6_{Vd} \overline{d}\gamma_\mu d)+ (\overline{e}\gamma^\mu\gamma_5 e) (C^6_{Au} \overline{u}\gamma_\mu\gamma_5 u) \nonumber\\
&+ D^\nu  \bigg(\overline{e}\gamma^\mu \gamma_5 e\bigg) D_\nu \bigg(\frac{C^{8}_{1u}}{v^2} \overline{u}\gamma_\mu u + \frac{C^{8}_{1d}}{v^2} \overline{d}\gamma_\mu d\bigg)+ D^\nu  \bigg(\overline{e}\gamma^\mu e\bigg) D_\nu \bigg(\frac{C^{8}_{2u}}{v^2} \overline{u}\gamma_\mu\gamma_5 u + \frac{C^{8}_{2d}}{v^2} \overline{d}\gamma_\mu\gamma_5 d\bigg)\nonumber\\
&+  D^\nu  \bigg(\overline{e}\gamma^\mu e\bigg) D_\nu \bigg(\frac{C^{8}_{Vu}}{v^2} \overline{u}\gamma_\mu u + \frac{C^{8}_{Vd}}{v^2} \overline{d}\gamma_\mu d\bigg)+ D^\nu  \bigg(\overline{e}\gamma^\mu\gamma_5 e\bigg) D_\nu \bigg(\frac{C^{8}_{Au}}{v^2} \overline{u}\gamma_\mu\gamma_5 u \bigg)\bigg].
\label{eq:lagrange}
\end{align}
We have extended this parameterization to include the type-1 dimension-8 extensions of the usual operators. We will refer to this as the PVES basis in this paper. The dimension-6 portion of this phenomenological operator basis can be mapped onto the usual dimension-6 SMEFT basis via the transformation
\begin{align}
 C^6_{1u} =&\;2(g_R^e-g_L^e) (g_R^u+g_L^u)+ \frac{v^2}{2\Lambda^2} \left\{-\left(C_{lq}^{(1)}-C_{lq}^{(3)}\right) + C_{eu} +C_{qe} -C_{lu}\right\}\nonumber\\
C^6_{2u} =&\; 2(g_R^e+g_L^e) (g_R^u-g_L^u)+ \frac{v^2}{2\Lambda^2} \left\{-\left(C_{lq}^{(1)}-C_{lq}^{(3)}\right) + C_{eu} -C_{qe} +C_{lu}\right\}\nonumber\\
C^6_{1d} =&\;2(g_R^e-g_L^e) (g_R^d+g_L^d)+  \frac{v^2}{2\Lambda^2} \left\{-\left(C_{lq}^{(1)}+C_{lq}^{(3)}\right) + C_{ed} +C_{qe} -C_{ld}\right\}\nonumber\\
 C^6_{2d} =&\;2(g_R^e+g_L^e) (g_R^d-g_L^d)+ \frac{v^2}{2\Lambda^2} \left\{-\left(C_{lq}^{(1)}+C_{lq}^{(3)}\right) + C_{ed} -C_{qe} +C_{ld}\right\}\nonumber\\
C^6_{Vu} =&\; 2(g_R^e+g_L^e) (g_R^u+g_L^u)+ \frac{v^2}{2\Lambda^2} \left\{\left(C_{lq}^{(1)}-C_{lq}^{(3)}\right) + C_{eu} +C_{qe} +C_{lu}\right\}\nonumber\\
 C^6_{Au} =&\; 2(g_R^e-g_L^e) (g_R^u-g_L^u)+  \frac{v^2}{2\Lambda^2} \left\{\left(C_{lq}^{(1)}-C_{lq}^{(3)}\right) + C_{eu} -C_{qe} -C_{lu}\right\}\nonumber\\
 C^6_{Vd} =&\;2(g_R^e+g_L^e) (g_R^d+g_L^d)+ \frac{v^2}{2\Lambda^2} \left\{\left(C_{lq}^{(1)}+C_{lq}^{(3)}\right) + C_{ed} +C_{qe} +C_{ld}\right\}.
\label{eq:translation}
\end{align}
The dimension-8 part of the phenomenological operator basis can be mapped through the same transformation with the dimension-6 SMEFT coefficients replaced by their dimension-8 counterparts in Table~\ref{tab:operators} and an additional factor of $\frac{v^2}{\Lambda^2}$. We additionally note that the coefficients of $\mathcal{L}_{PV}$ have non-zero SM values unlike the SMEFT coefficients, due to the inclusion of SM gauge-boson exchanges in their definitions. For the dimension-8 transformation the SM offset is scaled by $\frac{v^2}{M_Z^2}$. These shifts can be expressed in terms of the left and right-handed fermion couplings to the $Z$-boson. The leading-order values for these form factors follow the conventions in~\cite{Denner:1991kt} and amount to
\begin{align}
    g^f_L = I_3^f - Q_f s_W^2, \;\;\;\;\; g^f_R = -Q_f s_W^2, \;\;\;\;\; g_Z = \frac{e}{s_W c_W}.
\end{align}
Finally, we note that the axial-axial down-type operators with coefficients $C^6_{Ad}$ and $C^8_{Ad}$ are omitted from the Lagrangian in Eq.~(\ref{eq:lagrange}). We will see later that the use of this basis helps reveal experimental sensitivity to specific ultraviolet completions of the SMEFT that are obscure in the SMEFT basis for four-fermion Wilson coefficients.

The basis of dimension-6 semi-leptonic four-fermion SMEFT operators is built from $SU(2)$ doublets and singlets and consists of seven independent operators after electroweak symmetry breaking. Na\"ively one would expect eight operators making up $\mathcal{L}_{PV}$. This basis is over-complete since it is formed from fields after electroweak symmetry breaking, and we can eliminate one operator by making use of the underlying $SU(2)$ symmetry.

\section{Review of Drell-Yan and PVES Formulae}\label{sec:DYsolid}

In this section we review the formulae describing the Drell-Yan process and the parity-violating asymmetry parameter $A_{PV}$ in PVES. The review of the Drell-Yan cross sections closely follows Ref.~\cite{Boughezal:2020uwq}.
\subsection{Review of Drell-Yan}
We first present the cross section for the partonic Drell-Yan process $q + \overline{q}\rightarrow e^+ + e^-$. We decompose the differential cross section into three SM pieces stemming from photon and Z-boson exchange and their interference, two terms for interference between SM and SMEFT for each of the dimension-6 and dimension-8 operators respectively and one piece for the SMEFT dimension-6 squared term:
\begin{align}
\frac{d\sigma_{q\bar{q}}}{dm_{ll}^2 dY dc_{\theta}} =&\;  \frac{1}{32 \pi m_{ll}^2
  \hat{s}} f_q(x_1) f_{\bar{q}}(x_2) \left\{
  \frac{d\hat{\sigma}^{\gamma\gamma}_{q\bar{q}}}{dm_{ll}^2 dY dc_{\theta}}
  +  \frac{d\hat{\sigma}^{\gamma Z}_{q\bar{q}}}{dm_{ll}^2 dY dc_{\theta}}
  +  \frac{d\hat{\sigma}^{ZZ}_{q\bar{q}}}{dm_{ll}^2 dY dc_{\theta}}
  \right. \nonumber \\
 +& \left.
    \frac{d\hat{\sigma}^{\gamma SMEFT6}_{q\bar{q}}}{dm_{ll}^2 dY dc_{\theta}}
  +  \frac{d\hat{\sigma}^{Z SMEFT6}_{q\bar{q}}}{dm_{ll}^2 dY dc_{\theta}}+  \frac{d\hat{\sigma}^{\gamma SMEFT8}_{q\bar{q}}}{dm_{ll}^2 dY dc_{\theta}}+  \frac{d\hat{\sigma}^{Z SMEFT8}_{q\bar{q}}}{dm_{ll}^2 dY dc_{\theta}}+  \frac{d\hat{\sigma}^{SMEFT 6^2}_{q\bar{q}}}{dm_{ll}^2 dY dc_{\theta}}
\right\}.
\end{align}  
Here the $x_i$ are the partonic momentum fractions and $f_q(x)$ the parton distribution function describing the probability of finding a parton $q$ of momentum fraction x inside the proton. $m_{ll}$ is the invariant mass of the two final state leptons and $Y$ is its rapidity. Finally, $c_\theta$ is the cosine of the center of mass scattering angle of the negatively charged lepton. The hadronic cross section for the Drell-Yan process is derived by summing over all possible initial state quarks found inside the proton and integrating over their momentum fractions $x_1$ and $x_2$. The explicit expressions for the three terms making up the SM contribution to the differential cross section for initial state up-type quarks are given by
\begin{align}
 \frac{d\hat{\sigma}^{\gamma\gamma}_{u\bar{u}}}{dm_{ll}^2 dY dc_{\theta}}=\;& \frac{32 \pi^2 \alpha^2
                                Q_u^2}{3}\frac{\mht^2+\mhu^2}{\mhs^2},\nonumber\\
\frac{d\hat{\sigma}^{\gamma Z}_{u\bar{u}}}{dm_{ll}^2 dY dc_{\theta}} =\;& -\frac{8 \pi \alpha Q_u g_Z^2
                                }{3}\frac{(g_R^u
                            g_L^e+g_R^eg_L^u)\mht^2+(g_R^u
                            g_R^e+g_L^eg_L^u)\mhu^2}{\mhs (\mhs-M_Z^2)},
                            \nonumber \\
\frac{d\hat{\sigma}^{ZZ}_{u\bar{u}}}{dm_{ll}^2 dY dc_{\theta}} =&\; \frac{ g_Z^4
                                }{3}\frac{((g_R^u
                            g_L^e)^2+(g_R^eg_L^u)^2)\mht^2+((g_R^u
                            g_R^e)^2+(g_L^eg_L^u)^2)\mhu^2}{(\mhs-M_Z^2)^2}.
\label{eq:DYSMcs}
\end{align}  
The interference terms between the SM and the leading dimension-6 contributions is
\begin{align}                        
  \frac{d\hat{\sigma}^{\gamma SMEFT6}_{u\bar{u}}}{dm_{ll}^2 dY dc_{\theta}}=& -\frac{8 \pi \alpha Q_u 
                                }{3\Lambda^2}\frac{(C_{lu}+C_{qe})\mht^2+(C_{eu}+C_{lq}^{(1)}-C_{lq}^{(3)})\mhu^2}{\mhs },
                            \nonumber \\
 \frac{d\hat{\sigma}^{Z SMEFT6}_{u\bar{u}}}{dm_{ll}^2 dY dc_{\theta}} =\;& \frac{2g_Z^2
                                }{3\Lambda^2}\frac{(g_R^ug_L^eC_{lu}+g_R^eg_L^uC_{qe})\mht^2+(g_R^ug_R^eC_{eu}+g_L^ug_L^eC_{lq}^{(1)}-g_L^ug_L^eC_{lq}^{(3)})\mhu^2}{\mhs-M_Z^2}.
\label{eq:DYdim6cs}
\end{align}
The contributions stemming from interference of the SM with the corresponding dimension-8 terms, as well as the dimension-6 squared pieces, are
\begin{align}
\frac{d\hat{\sigma}^{\gamma SMEFT8}_{u\bar{u}}}{dm_{ll}^2 dY dc_{\theta}} =\;&-\frac{8 \pi \alpha Q_u}{3\Lambda^4}\bigg[\left(C^{(1)}_{l^2u^2D^2}+C^{(1)}_{q^2e^2D^2}\right)\mht^2 + \left(C_{e^2u^2D^2}^{(1)}+C_{l^2q^2D^2}^{(1)}-C_{l^2q^2D^2}^{(3)}\right)\mhu^2\bigg]\nonumber\\
 \frac{d\hat{\sigma}^{Z SMEFT8}_{u\bar{u}}}{dm_{ll}^2 dY dc_{\theta}} =\;&\frac{2g_Z^2}{3\Lambda^4}\frac{\mhs}{\mhs-M_Z^2}\bigg[\left(g_R^ug_L^eC^{(1)}_{l^2u^2D^2}+g_R^eg_L^uC^{(1)}_{q^2e^2D^2}\right)\mht^2 \nonumber\\
 & \hspace{5.5em}+ \left(g_R^ug_R^eC^{(1)}_{e^2u^2D^2}+g_L^ug_L^eC_{l^2q^2D^2}^{(1)}-g_L^ug_L^eC_{l^2q^2D^2}^{(3)}\right) \mhu^2
 \bigg]\nonumber\\
\frac{d\hat{\sigma}^{SMEFT6^2}_{u\bar{u}}}{dm_{ll}^2 dY dc_{\theta}}=\;&\frac{1}{3\Lambda^4} \bigg[(C_{lu}^2+C_{qe}^2) \hat{t}^2 + (C_{eu}^2 + (C_{lq}^{(1)} - C_{lq}^{(3)})^2) \hat{u}^2\bigg].      
\label{eq:DYdim8cs}                        
\end{align}
These depend on the standard Mandelstam variables $\hat{s}, \hat{t}$ and $\hat{u}$. The down-type contributions to the cross section can be found by replacing the indices and Wilson coefficients as follows:
\begin{gather}
Q_u\rightarrow Q_d \;\;\;\;\; g_{L/R}^u \rightarrow g_{L/R}^d \;\;\;\;\; C_{lq}^{(3)}\rightarrow -C_{lq}^{(3)} \;\;\;\;\; C_{lu} \rightarrow C_{ld} \;\;\;\;\;  C_{eu} \rightarrow C_{ed}\nonumber\\
C^{(1)}_{l^2u^2D^2}\rightarrow C^{(1)}_{l^2d^2D^2} \;\;\;\;\; C^{(1)}_{e^2u^2D^2}\rightarrow C^{(1)}_{e^2d^2D^2} \;\;\;\;\;
C^{(3)}_{l^2q^2D^2}\rightarrow -C^{(3)}_{l^2q^2D^2}.
\end{gather}
\subsection{Review of PVES}
The SoLID parity-violating deep-inelastic scattering (PVDIS) experiment will scatter a $12\,\textrm{GeV}$ polarized electron beam from polarized proton or deuteron targets~\cite{Solid1, Solid2, Solid3}. Similarly will P2 scatter a $155\,\textrm{MeV}$ electron beam elastically from hydrogen and carbon targets~\cite{Becker:2018ggl}. A goal of both experiments is to measure the parity-violating asymmetry 
\begin{align}
A_{PV} = \frac{\sigma_R -\sigma_L}{\sigma_R + \sigma_L},
\end{align}
where the $R$ and $L$ subscripts refer to incoming left- and right-handed electrons respectively. SoLID measurements will be taken in the kinematic range of $2\,\textrm{GeV}^2<Q^2<10\,\textrm{GeV}^2$ and $x>0.2$. The data taken using the proton target will be used to further constrain the $d/u$ ratio in the proton while the data taken with the deuteron target will be used to constrain new physics. P2 is dedicated to extracting a precise low-energy measurement of the electroweak mixing angle from $A_{PV}$. Since the electroweak mixing angle is predicted by the Standard Model, its precise measurement can be used to constrain new physics that shifts its value from the SM prediction.

In the limit $Q^2 \ll M_Z^2$ the asymmetry parameter can be written as~\cite{Hobbs:2008mm, Mantry:2010ki}
\begin{align}
A_{PV} = -\left(\frac{G_F Q^2}{4\sqrt{2} \pi \alpha}\right)^2\left[Y_1(x,y,Q^2) a_1 + Y_3(x,y,Q^2) a_3\right],
\end{align}
with
\begin{align}
a_1 = \frac{2\sum_q{e_q C_{1q}(q(x)+\overline{q}(x))}}{\sum_q{e_q^2(q(x)+\overline{q}(x))}} \;\;\;\;\;\;\; a_3 = \frac{2\sum_q{e_q C_{2q}(q(x)-\overline{q}(x))}}{\sum_q{e_q^2(q(x)+\overline{q}(x))}}.
\end{align}
This is under the assumption that the parity violating interactions are parameterized by the phenomenological Lagrangian in Eq.~(\ref{eq:lagrange}). The kinematic dependencies are encoded in the $Y_i$ functions:
\begin{align}
    Y_1(x,y,Q^2) =& \frac{1+(1-y)^2-y^2(1-r^2/(1+R^{\gamma Z}))-2xyM/E}{1+(1-y)^2-y^2(1-r^2/(1+R^\gamma))-2xyM/E}\left(\frac{1+R^{\gamma Z}}{1+R^\gamma}\right), \nonumber\\
     Y_3(x,y,Q^2) =& \frac{1-(1-y)^2}{1+(1-y)^2-y^2(1-r^2/(1+R^\gamma))-2xyM/E}\left(\frac{r^2}{1+R^\gamma}\right),
\end{align}
which in turn depend on $r = 1 +\frac{4M^2x^2}{Q^2}$, the electroweak structure functions $R^i = R^i(x,Q^2)$ as well as the target mass $M$ and beam energy $E$. These formulae are the basis for the projections of how well SoLID~\cite{Solid1} and P2~\cite{Becker:2018ggl} will be able to constrain the parity-violating coefficients $C_{iq}$. Due to the scalar nature of the Carbon nucleus used in the P2 experiment, the measurement will be insensitive to the $C_{2q}$ coefficients and will put bounds on $C_{1q}$. SoLID is expected to be more sensitive to $C_{2q}$.

We wish to emphasize at this point that it is important to analyze data from the LHC and PVES experiments in both the traditional SMEFT basis and also using the $C_{1q}$ and $C_{2q}$ coefficients instead. Since these coefficients map to a mix of SMEFT Wilson coefficients it is tempting to simply consider the resulting bounds in the SMEFT basis only. However, some ultraviolet completions, for example those with a leptophobic $Z'$~\cite{GonzalezAlonso:2012jb,Buckley:2012tc} only lead to a modification of the $C_{2q}$ while the $C_{1q}$ retain their SM values. The search for $Z'$ bosons is an ongoing effort at the LHC~\cite{Aaboud:2017buh} and it is important to be able to clearly interpret which regions of parameter space have been excluded by other experiments. While these results can be recovered in the SMEFT basis by choosing the appropriate combinations of Wilson coefficients, they are more easily seen in the basis consisting of the $C_{1q}$ and $C_{2q}$.

\section{Complementarity between LHC Drell-Yan data and PVES}\label{sec:Comparison}

In order to illustrate the impact that both SoLID and P2 have on breaking the degeneracies present in the Drell-Yan data we consider four example scenarios.\\
\begin{itemize}
    \item We first reproduce the SoLID projection of Ref.~\cite{Solid1} as well as the P2 projection of Ref.~\cite{Becker:2018ggl} in the PVES basis of Eq.~(\ref{eq:lagrange}). We study the constraints imposed by LHC Drell-Yan data in this parameter space and analyze the impact of combining the results of these experiments.
    
    \item We next consider the SMEFT basis and turn on only dimension-6 operators. We consider the combined fits of Wilson-coefficient combinations for which the Drell-Yan data has been previously shown to exhibit flat directions~\cite{Boughezal:2020uwq}, and study how SoLID and P2 can lift these degeneracies.
    
    \item We turn on both dimension-6 and dimension-8 operators in the basis of Eq.~(\ref{eq:lagrange}). In this basis it is easy to see that SoLID is uniquely capable of disentangling coefficients of type $C_{2q}$ while P2 probes coefficients of type $C_{1q}$. 
    
    \item Finally we turn on both dimension-6 and dimension-8 operators in the SMEFT basis.  We show that LHC Drell-Yan measurements can only poorly differentiate certain dimension-6 operators from dimension-8 effects, and then show that these degeneracies are lifted by PVES $A_{PV}$ measurements. 
    
\end{itemize}

For the analysis of the Drell-Yan process we follow Ref.\cite{Boughezal:2020uwq} and use the ATLAS data set presented in~\cite{Aad:2016zzw}, since it both goes to high invariant mass and is measured to high precision. We use the fully-correlated experimental errors. The SoLID collaboration anticipates using the data obtained from a hydrogen target to better determine the $d/u$ ratio inside the proton. Measurements taken with deuteron targets are dedicated to BSM searches. The SoLID projection~\cite{Solid1} is based on the assumption that the data available will stem from a single bin $0.4<x<0.5$ taken at $Q^2=6\,\textrm{GeV}^2$ with a $12\,\textrm{GeV}$ electron beam~\footnote{We thank K.~Kumar and P.~Souder for discussions on how to reproduce the SoLID projections.}. Measurements from the other available bins will serve to keep the systematic error under control. The total error for the measurement, including statistical uncertainties as well as both experimental and theoretical systematic errors, is predicted to be $0.6\%$. The P2 projection~\cite{Becker:2018ggl} assumes that the data stems from the scattering of a $150\,\textrm{MeV}$ polarized electron beam off hydrogen and carbon targets. It is assumed that the electroweak mixing angle extracted from these measurements will have a $0.3\%$ relative error, stemming mostly from the uncertainty of the beam polarization. This projection furthermore includes already determined constraints from SLAC~\cite{Anthony:2005pm}, atomic parity violation in Cesium~\cite{Wood:1997zq} as well as the anticipated bounds from the QWeak experiment~\cite{Androic:2017ppx}. 

To quantify the deviation from SM predictions we define a $\chi^2$ test statistic according to
\begin{align}
    \chi^2 = \sum_{i,j}^{\# \textrm{of observables}}{\frac{(\sigma_i^\textrm{SM}-\sigma_i^\textrm{SMEFT})(\sigma_j^\textrm{SM}-\sigma_j^\textrm{SMEFT})}{\Delta\sigma^2_{ij}}}.
\end{align}
For both the SoLID and P2 experiments there is no data available, yet so we project the SM to be correct and the SMEFT to measure simply the deviation from the SM allowed by the data. The sum over observables included all Drell-Yan invariant mass bins from Ref.~\cite{Aad:2016zzw} as well as the SoLID measurements and all observables making up the P2 projection.

We note that the Wilson coefficients are in principle dependent on the renormalization scheme chosen and are therefore subject to renormalization group running~\cite{Jenkins:2013zja,Jenkins:2013wua,Alonso:2013hga}. The measurements at LHC are taken at higher scales than the low-energy experiments. However, since we are performing an analysis to demonstrate the power of the PVES experiments before the availability of the actual data we neglect these effects here. They can be straightforwardly included once the data is available. 

\subsection{Dimension-6 analysis in the ${\cal L}_{PV}$ basis}
\begin{figure}[h!]
\centering
\includegraphics[width=.6\textwidth]{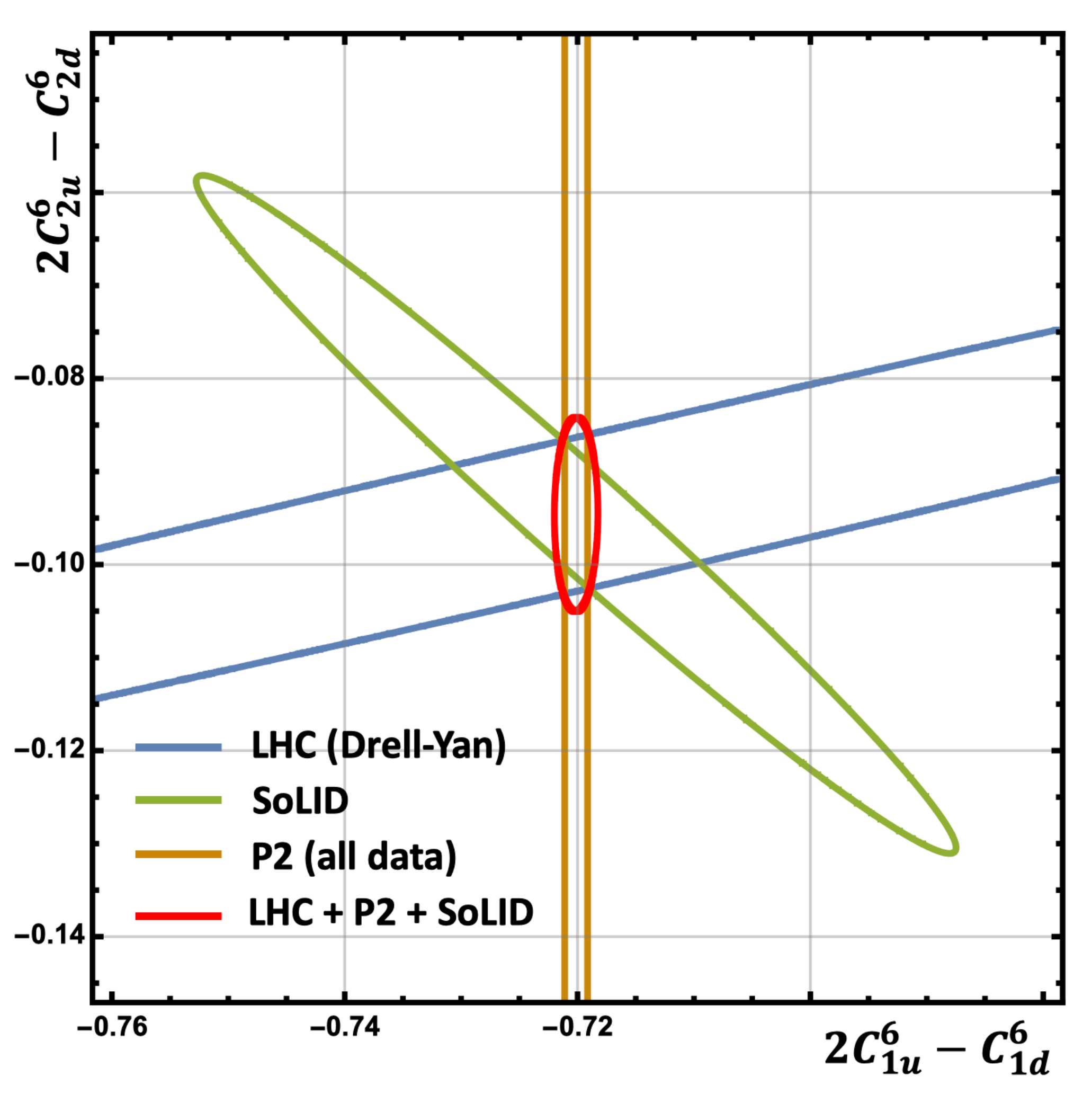}
\caption{Combination of the 68\% C.L. bounds derived from Drell-Yan data, P2 and SoLID in the dimension-6 $C_{iu}/C_{id}$ basis. Note the non-zero SM at the center of the ellipses. It corresponds to the loop-corrected first terms of Eq.~(\ref{eq:translation})~\cite{Akhundov:1985fc,Beenakker:1988pv}. In the case of P2 we include projections for data taken with both hydrogen and carbon targets and a projection for the QWeak experiment as well as available data from atomic-parity violation.}
\label{fig:Solidbasis}
\end{figure}

We begin by considering the constraints in the traditional basis used for PVES in Eq.~(\ref{eq:lagrange}). Both P2 and SoLID are designed to access different combinations of PVES coefficients. P2 will constrain $2C_{1u}^6 + C_{1d}^6$ through the hydrogen measurements as well as the combination $C_{1u}^6 + C_{1d}^6$ with the carbon target data. The anticipated bounds on the individual coefficients $C^6_{1i}$ are therefore expected to be tight enough to be neglected in the SoLID analysis. SoLID is thus expected to be able to focus on setting bounds on the combination $2C_{2u}^6 - C_{2d}^6$. To illustrate this point in our analysis, and compare with the available Drell-Yan data we set only the combinations of parameters $2C_{1u}-C_{1d}$ and $2C_{2u}-C_{2d}$ to be non-zero. The orthogonal combinations of parameters are assumed to vanish. The same combinations of parameters can be arranged for the Drell-Yan cross section using the conversion of Eq.~(\ref{eq:translation}). We note that all dimension-8 operators are turned off for this case. Similarly we neglect contributions from squared dimension-6 operators which are formally of the same order. Using the $\chi^2$ function defined above we study the constraints from SoLID, from P2, from LHC Drell-Yan data, and from the combination of all three experiments. As a check of our analysis procedure we note that our SoLID constraint ellipse matches that of Ref.~\cite{Solid1}. Our results are shown in Fig.~\ref{fig:Solidbasis}. The constraint ellipses associated with the SoLID and LHC measurements are nearly orthogonal, indicating that the directions poorly probed by Drell-Yan will be strongly constrained by SoLID. This is shown explicitly by the combined ellipse in Fig.~\ref{fig:Solidbasis}, which features a significantly smaller allowed parameter space. The P2 constraints are strong but can only probe the $C_{1q}$ direction.
\subsection{Dimension-6 analysis in the SMEFT basis} We now switch to the SMEFT basis. To illustrate the constraining power these low-energy experiments have in the space of SMEFT coefficients, we investigate example Wilson coefficient choices that lead to flat directions in the parameter space. The choice of these examples is adapted from Ref.~\cite{Boughezal:2020uwq}.
\subsubsection{Dimension-6: Case 1}
\begin{figure}[h!]
\centering
\includegraphics[width=0.68\textwidth]{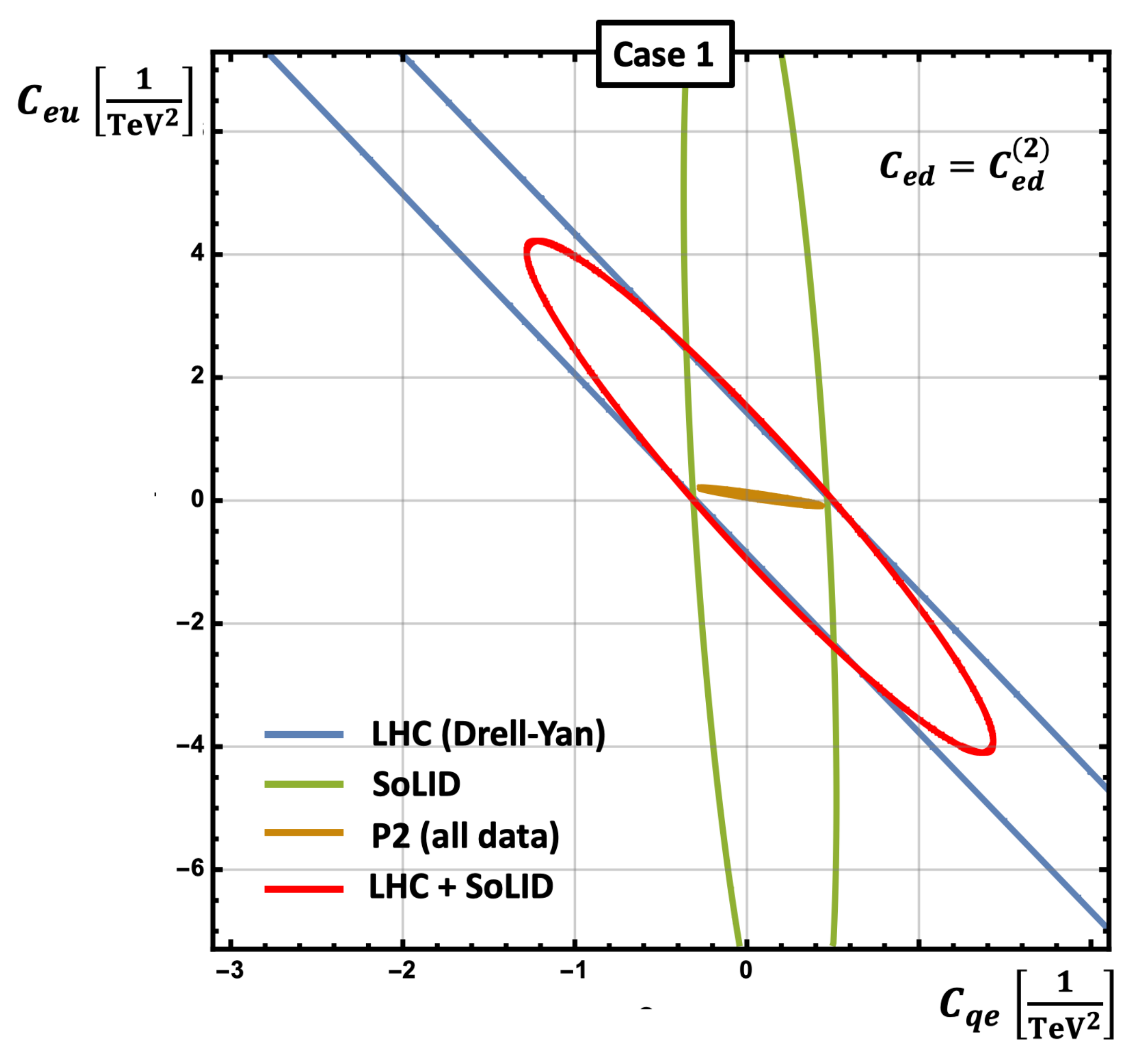}
\caption{Combining the 68\% C.L. bounds derived from Drell-Yan data for the Case 1 scenario combined with the projections for P2 and SoLID in the SMEFT basis. The operators are normalized to $\Lambda = 3\,\textrm{TeV}$.}
\label{fig:Case1}
\end{figure}

 In a first scenario we assume $C_{eu}, C_{qe}$ and $C_{ed}$ to be non-zero and truncate all Matrix elements at order $\frac{1}{\Lambda^2}$. We see from Eq.~(\ref{eq:DYdim6cs}) that these coefficients appear in both terms containing $\hat{t}$ as well as terms proportional to $\hat{u}$. In principle there should not be a flat direction. This is however not the case when analyzing the currently-available high-invariant mass LHC data~\cite{Boughezal:2020uwq}; after performing the angular integrations relevant for the $m_{ll}$ distributions the discriminatory power in the angular distributions vanishes.
After performing the angular integration we find that the SMEFT contributions vanishes for 
\begin{align}
    C_{ed} = \frac{Q_u e^2 - g_Z^2g_L^u g_R^e}{Q_u e^2 - g_Z^2 g_R^e g_R^u}\frac{Q_d e^2 - g_Z^2 g_R^e g_R^d}{Q_de^2-g_Z^2g_L^d g_R^e}C_{eu} \equiv C_{ed}^{(2)}.
\end{align}
 We perform a 2-dimensional $\chi^2$ fit after projecting $C_{ed}$ down to $C_{ed}^{(2)}$. The constraints are shown in Figure~\ref{fig:Case1}. Due to the flat direction the constraints derived from Drell-Yan data are fairly loose and only constrain the absolute values of $C_{eu}$ and $C_{qe}$ to be smaller than about $15$ and $40$ respectively (we normalize the operators to $\Lambda = 3\,\textrm{TeV}$ and limit the plot to the relevant region in which the contours intersect). Including projected SoLID data removes this flat direction since PVES is proportional to different combinations of SMEFT parameters. With both LHC and SoLID the 
same coefficients are constrained to about $1$ and $4$.

The P2 experiment will be able to constrain this combination of parameters even further. P2, shown by the brown ellipse in Fig.\ref{fig:Case1}, constrains $|C_{qe}|$ to smaller than about 0.2 and $|C_{eu}|$ to smaller than 0.4. This serves to illustrate an important point that generically appears when analyzing PVES constraints in the SMEFT basis. The $C_{1q}$ coefficients are linear combinations of all relevant SMEFT coefficients, as evident from Eq.~(\ref{eq:translation}). At the same time the anticipated constraints from the P2 experiments on $C_{1q}$ are extremely strong, with an expected precision on certain linear combinations of the $C_{1q}$ approaching a few times 0.001~\cite{Becker:2018ggl}. In contrast the expected SoLID constraints on the $C_{2q}$ are on the order of a few times 0.01. Since a generic SMEFT Wilson coefficient projects onto both $C_{1q}$ and $C_{2q}$ the constraints from P2 will be far stronger than those from SoLID on those coefficients. The power of the SoLID experiment is only revealed when certain linear combinations of SMEFT coefficients are considered, such as those motivated by leptophobic $Z'$ ultraviolet completions~\cite{GonzalezAlonso:2012jb,Buckley:2012tc}. Although the physics is basis independent, results are sometimes easier to see in a certain basis, and this is a strong motivation to consider constraints in both the SMEFT and PVES bases when analyzing these experiments. 

\subsubsection{Dimension-6: Case 2}

\begin{figure}[h!]
\centering
\includegraphics[width=0.68 \textwidth]{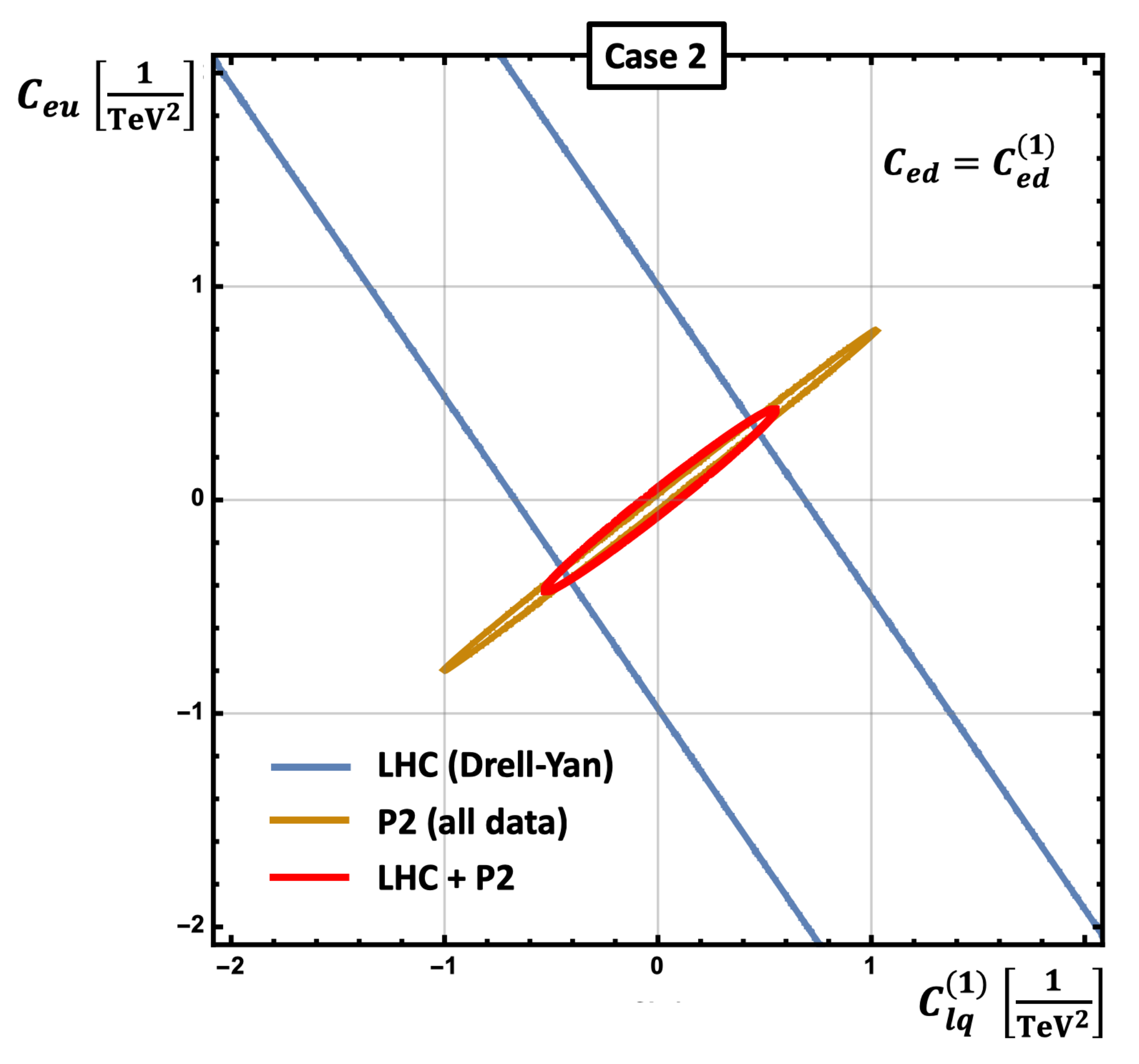}
\caption{Combining the 68\% C.L. bounds derived from Drell-Yan data for the Case 2 scenario combined with the projections for SoLID and P2 in the SMEFT basis. The operators are normalized to $\Lambda = 3\,\textrm{TeV}$.}
\label{fig:Case2}
\end{figure}

 We next assume that $C_{lq}^{(1)}, C_{eu}$ and $C_{ed}$ are non-zero. As discussed in detail in Ref.~\cite{Boughezal:2020uwq} these coefficients only contribute to the $\hat{u}^2$ term in Eq.~(\ref{eq:DYdim6cs}). In the high-invariant mass limit we can arrange these coefficients so that the correction to the Drell-Yan cross section vanishes. The particular combination of dimension-6 SMEFT contributions that leads to this is~\cite{Boughezal:2020uwq}
\begin{align}
    C_{ed} = \frac{Q_u e^2 - g_Z^2g_R^u g_R^e}{Q_u e^2 - g_Z^2 g_L^e g_L^u}\frac{Q_d e^2 - g_Z^2 g_L^e g_L^d}{Q_de^2-g_Z^2g_R^d g_R^e}C_{eu} \equiv C_{ed}^{(1)},
\end{align}
indicating a flat direction in the dimension-6 parameter space. We again illustrate the bounds the Drell-Yan data can set with a two-dimensional fit in Fig.~\ref{fig:Case2} where we project down onto the flat direction $C_{ed} = C_{ed}^{(1)}$. We contrast the LHC bounds with the projected P2 bounds. As discussed in the previous section the expected constraints from SoLID are weaker, and we do not show them explicitly here. The Drell-Yan analysis only manages to constrain the absolute values of $C_{lq}^{(1)}$ and $C_{eu}$ to be smaller than $30$ and $40$ respectively. Combining the two measurements reduces the allowed dimension-6 parameter space considerably. At $68\%$ confidence we can constrain both Wilson coefficients well below unity.

\subsubsection{Dimension-6: Case 3}

Finally we consider one more combination of dimension-6 SMEFT coefficients. In order to explore a larger swath of the dimension-6 parameter space we turn on the up-type operators $C_{lu}$ and $C_{eu}$. This serves to illustrate a scenario in which probes from all three experiments become comparable. In this example we can also show the potential of combining the experiments even if the observables do not suffer from flat directions, which is the case for this choice of Wilson coefficients.

\begin{figure}[h!]
\centering
\includegraphics[width=0.68\textwidth]{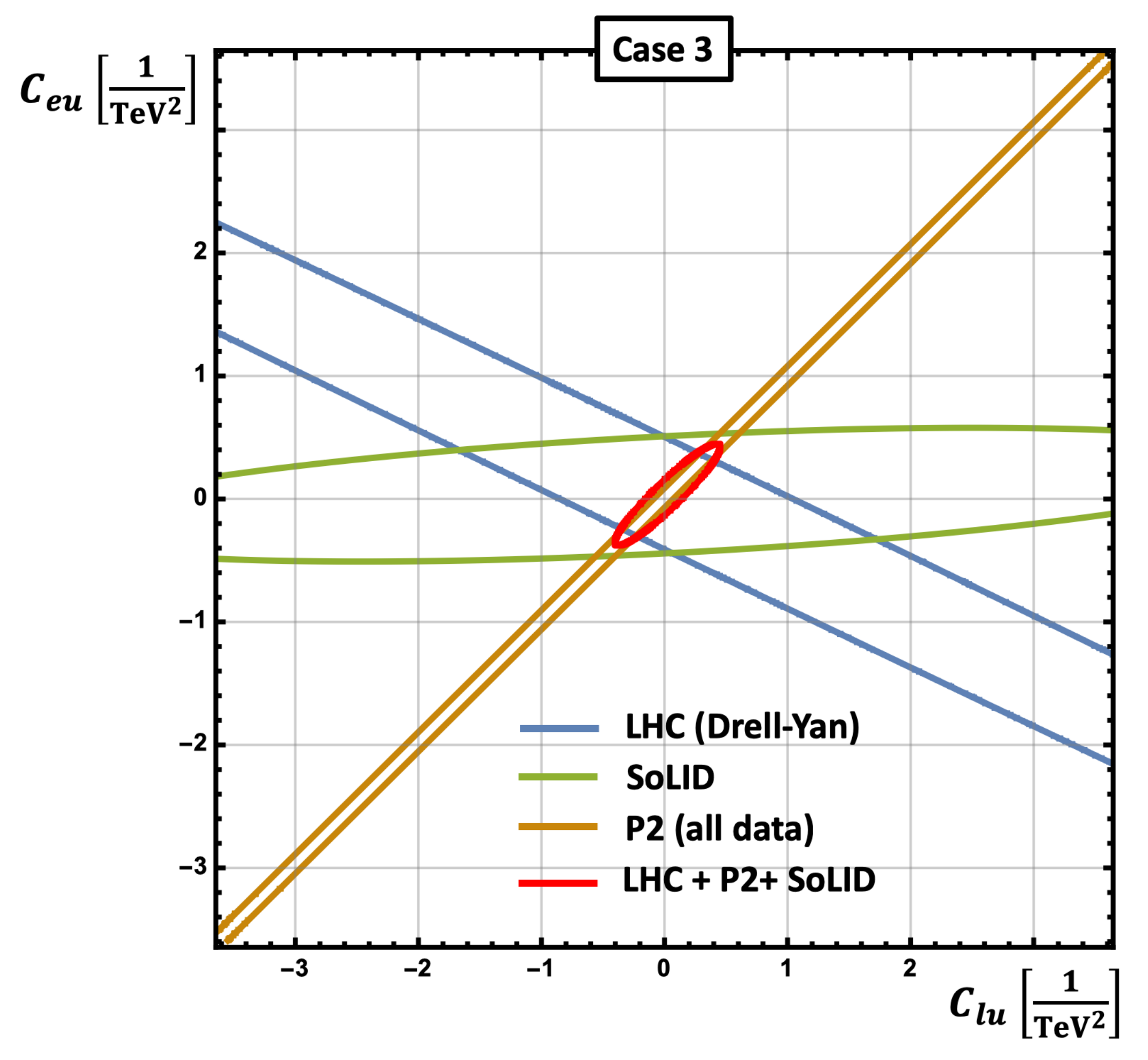}
\caption{Combining the 68\% C.L. bounds derived from Drell-Yan data for the Case 3 scenario combined with the projections for P2 and SoLID in the SMEFT basis. The operators are normalized to $\Lambda = 3\,\textrm{TeV}$.}
\label{fig:Case3}
\end{figure}

After performing the 2-dimensional $\chi^2$ fit we arrive at bounds shown in Fig.~\ref{fig:Case3}. A fit to the Drell-Yan data is able to constrain $|C_{lu}|$ and $|C_{eu}|$ to be smaller than about $15$ and $10$ respectively. SoLID constrains the same coefficients to $|C_{lu}|< 5$ and $|C_{eu}|< 0.5$. P2 exhibits a flat direction and can only constrain the combination $|C_{lu} - C_{eu}|$ to be smaller than $0.08$. Combining all three measurements enables us to dramatically cut down on the available parameter space and allows us to constrain the absolute value of each individual coefficient to be smaller than $0.5$.
\subsection{Joint dimension-6 and dimension-8 fits in the ${\cal L}_{PV}$ basis}
\begin{figure}[h!]
\centering
\includegraphics[width=0.6 \textwidth]{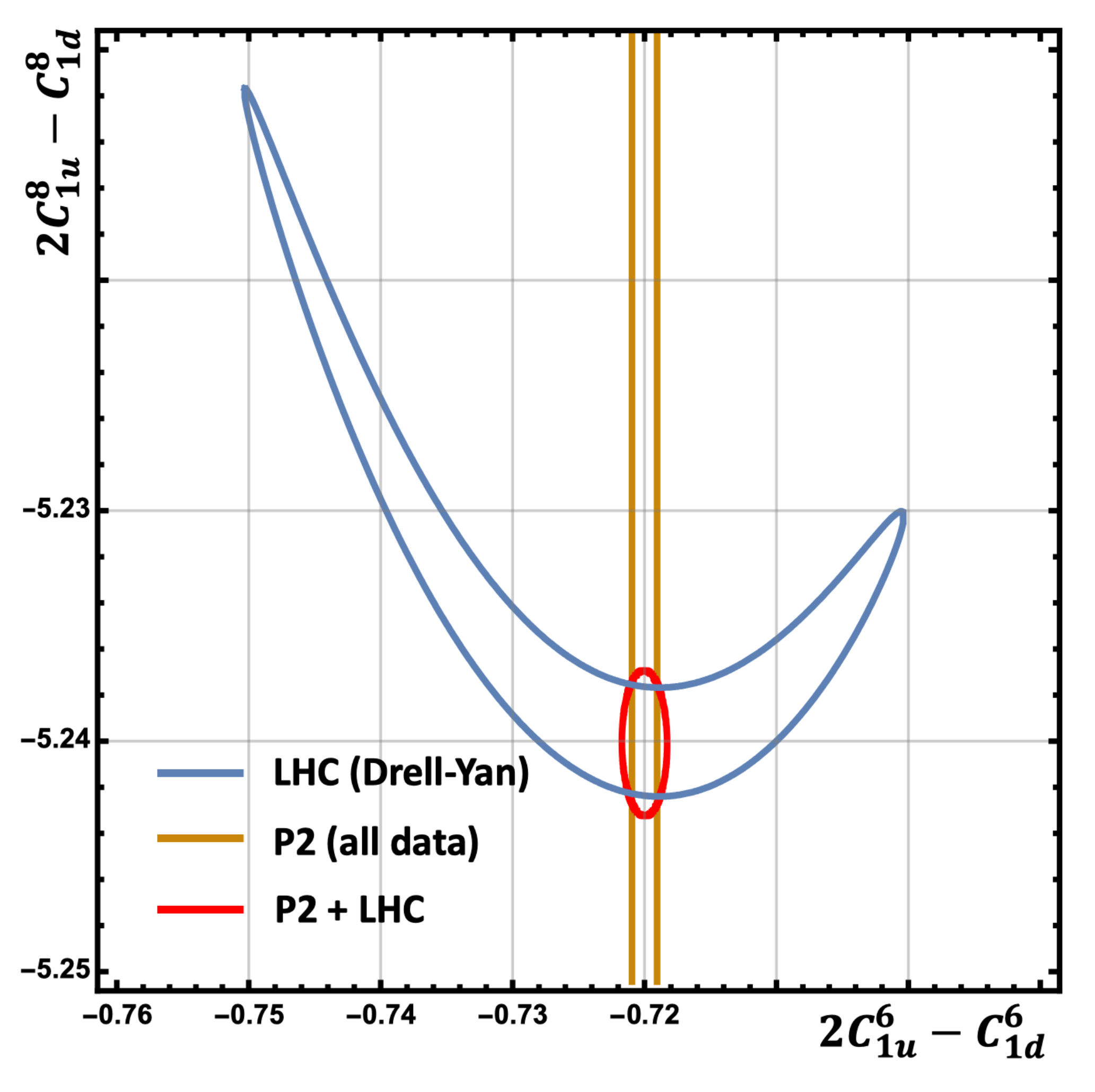}
\caption{Combining the 68\% C.L. bounds derived from Drell-Yan data and the P2 projection in the $C_{iu}/C_{id}$ basis contrasting dimension-6 and dimension-8 contributions.}
\label{fig:PVdim8P2}
\end{figure}

We now study the effect of simultaneously turning on both dimension-6 and dimension-8 operators. For consistency of the EFT expansion we also include the square of dimension-6 coefficients in our calculations. Naively we would expect that measurements of the invariant mass distribution at the LHC would distinguish between dimension-6 and dimension-8 effects since they scale differently with energy. However, we find significant interplay between the dimension-6 squared effects and the dimension-8 operators, leading to degeneracies in the parameter space. Both SoLID and P2 help to resolve these degeneracies. As low-energy experiments, they have completely negligible dependence on ${\cal O}(1/\Lambda^4)$ effects, and are therefore sensitive to only ${\cal O}(1/\Lambda^2)$ dimension-6 effects. To illustrate how SoLID and P2 complement each other in constraining the parameter space we first perform an analysis in the the PVES basis defined by Eq.~\ref{eq:lagrange}. We begin by turning on the combination $2C_{1u} - C_{1d}$ and its dimension-8 extension $2C^8_{1u} - C^8_{1d}$ for both P2 and Drell-Yan at LHC. As mentioned previously the P2 probes of the $C_{1q}$ coefficients are stronger than the SoLID probes of these terms. We limit ourselves to the two-derivative extension discussed in Section~\ref{sec:smeft}. The placement of the derivatives leads to an overall energy-dependent scaling of the dimension-6 term. Performing a fit to the LHC data leads to the blue contour in Fig.~\ref{fig:PVdim8P2}. The contour is no longer a simple ellipse due to the presence of dimension-6 squared terms. The $1\sigma$ contour for P2 is shown in brown and does not exhibit a dependence on the dimension-8 operators, as expected. P2 significantly reduces the parameter space allowed by the LHC Drell-Yan data. The elongation of the LHC bounds occurs due to cancellations between dimension-8 effects and dimension-6 squared terms. These can be removed by P2 measurements.

\begin{figure}[h!]
\centering
\includegraphics[width=0.6 \textwidth]{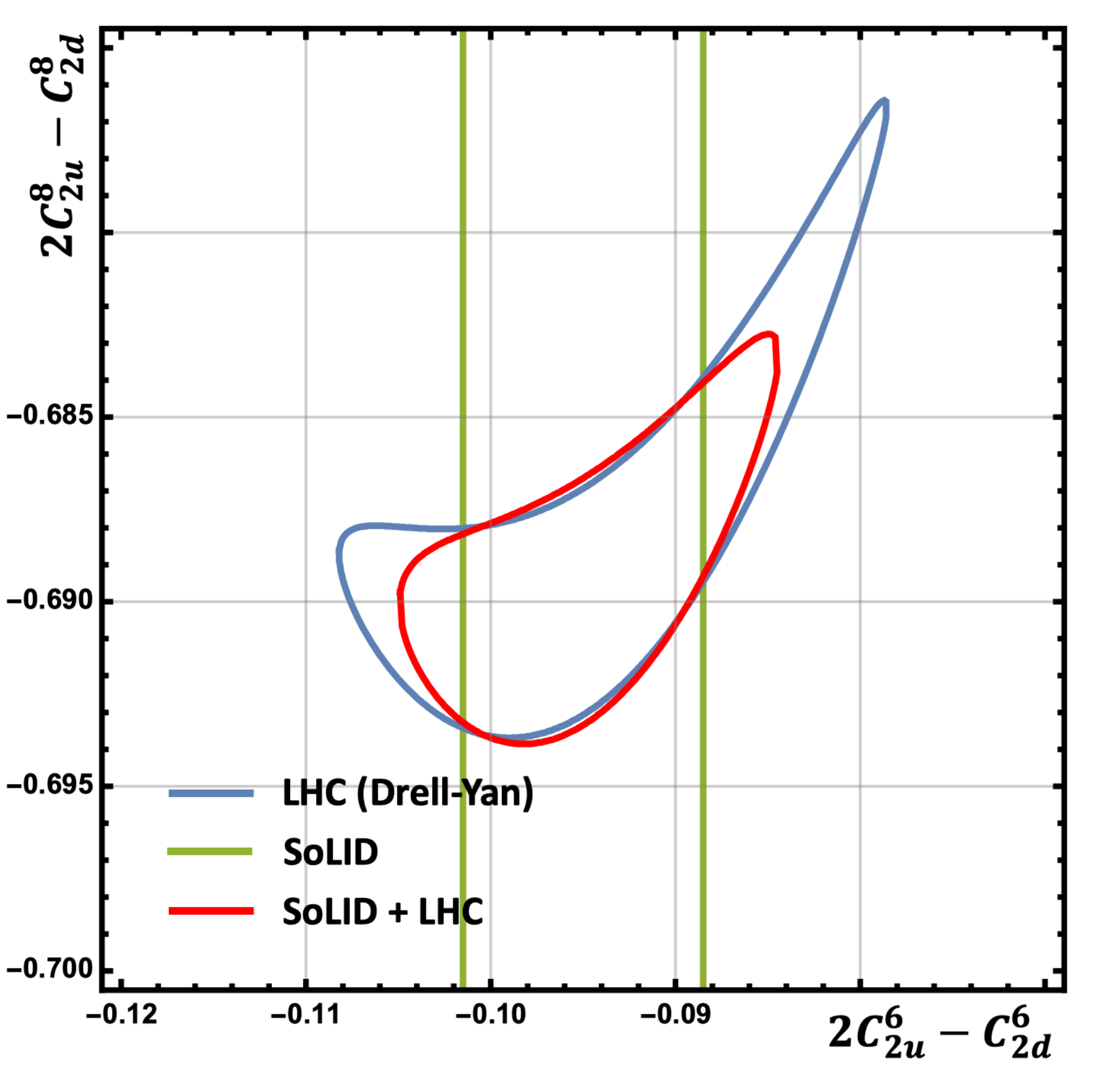}
\caption{Combining the 68\% C.L. bounds derived from Drell-Yan data and the SoLID projection in the $C_{iu}/C_{id}$ basis contrasting dimension-6 and dimension-8 contributions.}
\label{fig:PVdim8Solid}
\end{figure}

We can probe the other scenario where instead the $C_{2q}$ coefficients are turned on, and the SoLID probes are more important than P2 ones. We consider the scenario where $2C^6_{2u} - C^6_{2d}$ and its dimension-8 extension $2C^8_{2u} - C^8_{2d}$ are turned on. The resulting bounds for Drell-Yan at LHC and SoLID in terms of these operators are shown in Fig.~\ref{fig:PVdim8Solid}. Again, the low-energy PVDIS experiment SoLID can rule out parameter space allowed by the LHC, although the effects are not as significant as in the previous P2 analysis.

\subsection{Joint dimension-6 and dimension-8 fits in the SMEFT basis}

To illustrate how the inclusion of low-energy PVES experiments impacts the bounds on SMEFT coefficients we now translate the fits of the previous section into the SMEFT basis. As an example we consider the type-1 dimension-8 operator $C_{l^2q^2D^2}^{(1)}$, which we see from Table~\ref{tab:operators} is a two-derivative extension of the dimension-6 term $C_{lq}^{(1)}$. The stretching of the constraint contour occurs because the SMEFT modification of the Drell-Yan cross section vanishes for the combination
\begin{align}
  C^{(1)}_{l^2q^2D^2} =  C_{lq}^{(1)}\left[\frac{1}{2}\frac{C_{lq}^{(1)}}{e^2 Q_u- g_z^2g_L^e g_L^u} - \frac{\Lambda^2}{\hat{s}}\right]
\end{align}
in the high-energy limit $s \gg M_Z^2$. Although this condition changes as the invariant mass bin changes, most sensitivity comes from the higher invariant mass bins, leading to the long tails seen in the plot. The asymmetry parameter $A_{PV}$ is in principle dependent on the same linear combination of coefficients. However, the dimension-8 piece is suppressed by $Q^2/\Lambda^2$, and the P2 projection is therefore largely independent of $C_{l^2q^2D^2}^{(1)}$. Combining the Drell-Yan bounds with the projected P2 results constrains $|C_{lq}^{(1)}|$ to be less than $0.1$, while $|C_{l^2q^2D^2}^{(1)}|$ is bound to be smaller than $8$.
\begin{figure}[h!]
\centering
\includegraphics[width=0.71 \textwidth]{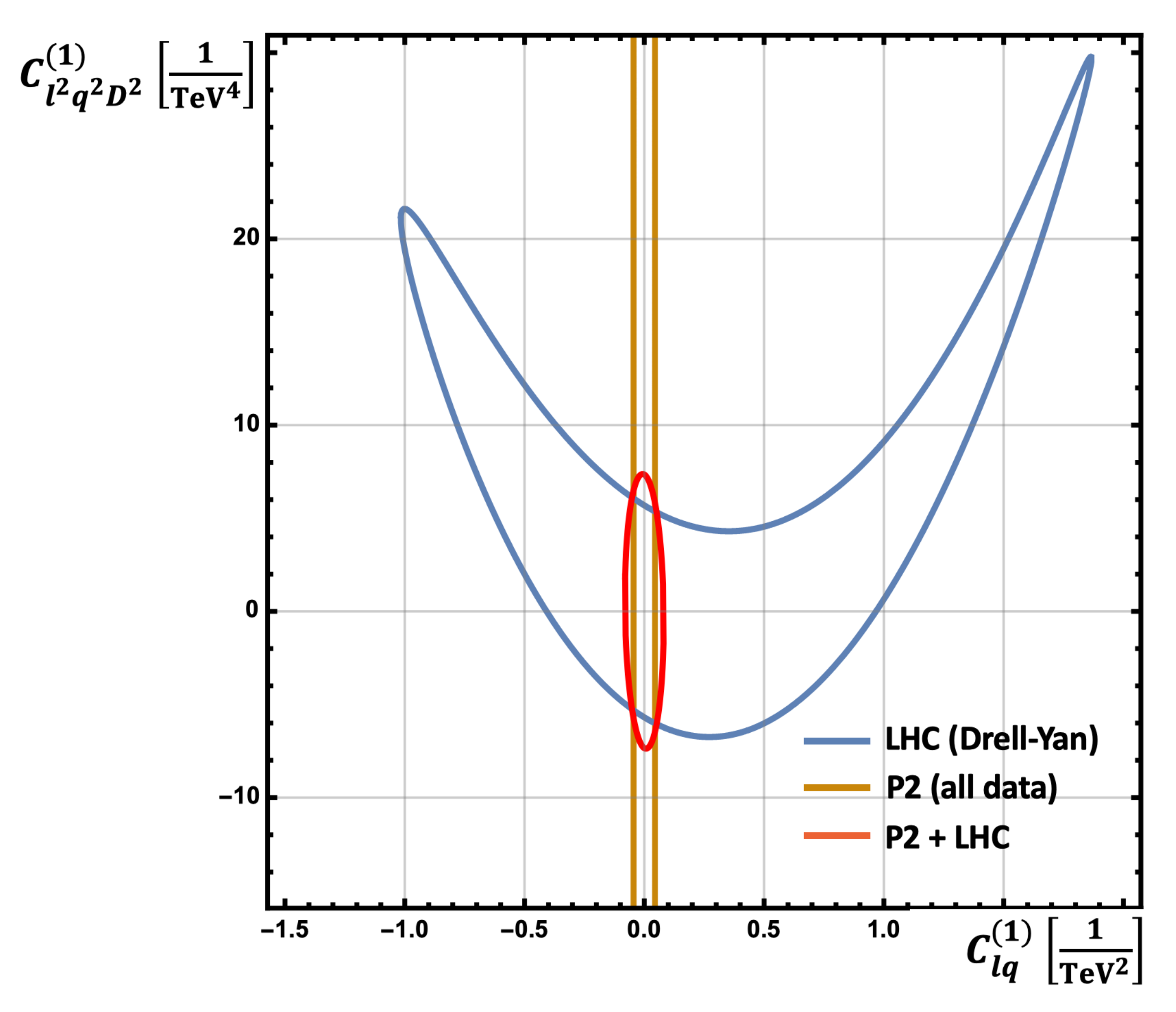}
\caption{Example SMEFT plot combining the 68\% C.L. bounds derived from Drell-Yan data and the P2 projection for $C_{lq}^{(1)}$ and its type-1 dimension-8 extension $C_{l^2q^2D^2}^{(1)}$. The operators are normalized to $\Lambda = 3\,\textrm{TeV}$.}
\label{fig:SMEFTdim8P2}
\end{figure}
%
\section{Conclusions}  \label{sec:conc}
In this work we have studied the potential impact that future PVES experiments SoLID and P2 will have on disentangling degeneracies in SMEFT fits, and in separating dimension-6 from dimension-8 effects. Both experiments can discriminate between combinations of dimension-6 operators that cannot be resolved by existing Drell-Yan data at the LHC. We have studied several such examples motivated by previous work~\cite{Boughezal:2020uwq} to demonstrate this point. A generic issue that we have discussed extensively in this work is the importance of studying Wilson-coefficient bases motivated by specific ultraviolet examples in order to properly evaluate the impact of different experiments. In the situation here the use of the traditional PVES basis in terms of $C_{1q}$ and $C_{2q}$ illustrates complementarity between the SoLID and P2 experiments difficult to see in the SMEFT basis. We have illustrated through the use of the PVES basis that the bounds on parameter space set by SoLID and P2 are complementary and how these bounds translate into the standard SMEFT basis. We have also emphasized that the lower energies of the SoLID experiment can be exploited to separate dimension-6 from dimension-8 effects when combined with high invariant-mass LHC data. To demonstrate this point we have presented example fits containing both dimension-6 four-fermion operators and their dimension-8 extensions. Combined fits of LHC data and projected SoLID and P2 data break degeneracies between dimension-6 and dimension-8 effects and tighten bounds on individual Wilson coefficients considerably.
\section*{Acknowledgments}
We thank K.~Kumar and P.~Souder for motivating this work and for helpful discussions. R.~B. is supported by the DOE contract DE-AC02-06CH11357. F.~P. and D.~W. are supported by the DOE grants DE-FG02-91ER40684 and DE-AC02-06CH11357. The U.S. Government retains for itself, and others acting on its
behalf, a paid-up nonexclusive, irrevocable worldwide license in said article to reproduce,
prepare derivative works, distribute copies to the public, and perform publicly and display
publicly, by or on behalf of the Government.


\begin{thebibliography}{99}
\frenchspacing

\bibitem{Buchmuller:1985jz} 
  W.~Buchmuller and D.~Wyler,
  Nucl.\ Phys.\ B {\bf 268}, 621 (1986).
  doi:10.1016/0550-3213(86)90262-2

\bibitem{Arzt:1994gp}
C.~Arzt, M.~B.~Einhorn and J.~Wudka,
Nucl. Phys. B \textbf{433}, 41-66 (1995)
doi:10.1016/0550-3213(94)00336-D
[arXiv:hep-ph/9405214 [hep-ph]].

\bibitem{Grzadkowski:2010es} 
  B.~Grzadkowski, M.~Iskrzynski, M.~Misiak and J.~Rosiek,
  JHEP {\bf 1010}, 085 (2010)
  doi:10.1007/JHEP10(2010)085
  [arXiv:1008.4884 [hep-ph]].
  
  \bibitem{Murphy:2020rsh}
C.~W.~Murphy,
JHEP \textbf{10}, 174 (2020)
doi:10.1007/JHEP10(2020)174
[arXiv:2005.00059 [hep-ph]].

\bibitem{Li:2020gnx}
H.~L.~Li, Z.~Ren, J.~Shu, M.~L.~Xiao, J.~H.~Yu and Y.~H.~Zheng,
[arXiv:2005.00008 [hep-ph]].



\bibitem{Han:2004az} 
  Z.~Han and W.~Skiba,
  Phys.\ Rev.\ D {\bf 71}, 075009 (2005)
  doi:10.1103/PhysRevD.71.075009
  [hep-ph/0412166].

\bibitem{Chen:2013kfa} 
  C.~Y.~Chen, S.~Dawson and C.~Zhang,
  Phys.\ Rev.\ D {\bf 89}, no. 1, 015016 (2014)
  doi:10.1103/PhysRevD.89.015016
  [arXiv:1311.3107 [hep-ph]].

\bibitem{Ellis:2014dva} 
  J.~Ellis, V.~Sanz and T.~You,
  JHEP {\bf 1407}, 036 (2014)
  doi:10.1007/JHEP07(2014)036
  [arXiv:1404.3667 [hep-ph]].
  
\bibitem{Wells:2014pga} 
  J.~D.~Wells and Z.~Zhang,
  Phys.\ Rev.\ D {\bf 90}, no. 3, 033006 (2014)
  doi:10.1103/PhysRevD.90.033006
  [arXiv:1406.6070 [hep-ph]].

 \bibitem{Falkowski:2014tna} 
  A.~Falkowski and F.~Riva,
  JHEP {\bf 1502}, 039 (2015)
  doi:10.1007/JHEP02(2015)039
  [arXiv:1411.0669 [hep-ph]].
  
  \bibitem{Cirigliano:2016nyn} 
  V.~Cirigliano, W.~Dekens, J.~de Vries and E.~Mereghetti
  Phys.\ Rev.\ D {\bf 94}, no. 3, 034031 (2016)
  doi:10.1103/PhysRevD.94.034031
  [arXiv:1605.04311 [hep-ph]].
  
  \bibitem{deBlas:2016ojx} 
  J.~de Blas, M.~Ciuchini, E.~Franco, S.~Mishima, M.~Pierini, L.~Reina and L.~Silvestrini,
  JHEP {\bf 1612}, 135 (2016)
  doi:10.1007/JHEP12(2016)135
  [arXiv:1608.01509 [hep-ph]].
  
  \bibitem{Hartmann:2016pil}
C.~Hartmann, W.~Shepherd and M.~Trott,
JHEP \textbf{03}, 060 (2017)
doi:10.1007/JHEP03(2017)060
[arXiv:1611.09879 [hep-ph]].
 
  \bibitem{Falkowski:2017pss} 
  A.~Falkowski, M.~González-Alonso and K.~Mimouni,
  JHEP {\bf 1708}, 123 (2017)
  doi:10.1007/JHEP08(2017)123
  [arXiv:1706.03783 [hep-ph]].
  
\bibitem{Biekotter:2018rhp} 
  A.~Biekoetter, T.~Corbett and T.~Plehn,
  SciPost Phys.\  {\bf 6}, no. 6, 064 (2019)
  doi:10.21468/SciPostPhys.6.6.064
  [arXiv:1812.07587 [hep-ph]].
  
\bibitem{Grojean:2018dqj}
C.~Grojean, M.~Montull and M.~Riembau,
JHEP \textbf{03}, 020 (2019)
doi:10.1007/JHEP03(2019)020
[arXiv:1810.05149 [hep-ph]].

\bibitem{Pomarol:2013zra} 
  A.~Pomarol and F.~Riva,
  JHEP {\bf 1401}, 151 (2014)
  doi:10.1007/JHEP01(2014)151
  [arXiv:1308.2803 [hep-ph]].
  
  \bibitem{DiVita:2017eyz}
S.~Di Vita, C.~Grojean, G.~Panico, M.~Riembau and T.~Vantalon,
JHEP \textbf{09}, 069 (2017)
doi:10.1007/JHEP09(2017)069
[arXiv:1704.01953 [hep-ph]].

\bibitem{Almeida:2018cld}
E.~da Silva Almeida, A.~Alves, N.~Rosa Agostinho, O.~J.~P.~\'Eboli and M.~C.~Gonzalez-Garcia,
Phys. Rev. D \textbf{99}, no.3, 033001 (2019)
doi:10.1103/PhysRevD.99.033001
[arXiv:1812.01009 [hep-ph]].
  
  \bibitem{Ellis:2018gqa}
J.~Ellis, C.~W.~Murphy, V.~Sanz and T.~You,
JHEP \textbf{06}, 146 (2018)
doi:10.1007/JHEP06(2018)146
[arXiv:1803.03252 [hep-ph]].
  
  \bibitem{Hartland:2019bjb} 
  N.~P.~Hartland, F.~Maltoni, E.~R.~Nocera, J.~Rojo, E.~Slade, E.~Vryonidou and C.~Zhang,
  JHEP {\bf 1904}, 100 (2019)
  doi:10.1007/JHEP04(2019)100
  [arXiv:1901.05965 [hep-ph]].

\bibitem{Brivio:2019ius} 
  I.~Brivio, S.~Bruggisser, F.~Maltoni, R.~Moutafis, T.~Plehn, E.~Vryonidou, S.~Westhoff and C.~Zhang,
  JHEP {\bf 2002}, 131 (2020)
  doi:10.1007/JHEP02(2020)131
  [arXiv:1910.03606 [hep-ph]].
  
  \bibitem{vanBeek:2019evb}
S.~van Beek, E.~R.~Nocera, J.~Rojo and E.~Slade,
SciPost Phys. \textbf{7}, no.5, 070 (2019)
doi:10.21468/SciPostPhys.7.5.070
[arXiv:1906.05296 [hep-ph]].

\bibitem{Aoude:2020dwv}
R.~Aoude, T.~Hurth, S.~Renner and W.~Shepherd,
[arXiv:2003.05432 [hep-ph]].


  \bibitem{Ellis:2020unq}
  J.~Ellis, M.~Madigan, K.~Mimasu, V.~Sanz and T.~You,
  [arXiv:2012.02779 [hep-ph]].
  
  
  \bibitem{Contino:2016jqw}
R.~Contino, A.~Falkowski, F.~Goertz, C.~Grojean and F.~Riva,
JHEP \textbf{07}, 144 (2016)
doi:10.1007/JHEP07(2016)144
[arXiv:1604.06444 [hep-ph]].

\bibitem{Liu:2016idz}
D.~Liu, A.~Pomarol, R.~Rattazzi and F.~Riva,
JHEP \textbf{11}, 141 (2016)
doi:10.1007/JHEP11(2016)141
[arXiv:1603.03064 [hep-ph]].
  

\bibitem{Degrande:2013kka}
C.~Degrande,
JHEP \textbf{02}, 101 (2014)
doi:10.1007/JHEP02(2014)101
[arXiv:1308.6323 [hep-ph]].

\bibitem{Hays:2018zze}
C.~Hays, A.~Martin, V.~Sanz and J.~Setford,
JHEP \textbf{02}, 123 (2019)
doi:10.1007/JHEP02(2019)123
[arXiv:1808.00442 [hep-ph]].

\bibitem{Bellazzini:2018paj}
B.~Bellazzini and F.~Riva,
Phys. Rev. D \textbf{98}, no.9, 095021 (2018)
doi:10.1103/PhysRevD.98.095021
[arXiv:1806.09640 [hep-ph]].

\bibitem{Ellis:2019zex}
J.~Ellis, S.-F.~Ge, H.-J.~He, R.-Q.~Xiao~Rui
Sci. China Phys. Mech. Astron. 64 (2021), no. 2 221062,
doi: 10.1088/1674-1137/44/6/063106 
[arXiv:2008.04298 [hep-ph]].

  
  \bibitem{Alioli:2020kez}
  S.~Alioli, R.~Boughezal, E.~Mereghetti and F.~Petriello,
  Phys. Lett. B \textbf{809}, 135703 (2020)
  doi:10.1016/j.physletb.2020.135703
  [arXiv:2003.11615 [hep-ph]].

\bibitem{Murphy:2020cly}
C.~W.~Murphy,
[arXiv:2012.13291 [hep-ph]].

\bibitem{Hays:2020scx}
C.~Hays, A.~Helset, A.~Martin and M.~Trott,
JHEP \textbf{11}, 087 (2020)
doi:10.1007/JHEP11(2020)087
[arXiv:2007.00565 [hep-ph]].

\bibitem{Ellis:2020ljj}
J.~Ellis, H.~J.~He and R.~Q.~Xiao,
Sci. China Phys. Mech. Astron. \textbf{64}, no.2, 221062 (2021)
doi:10.1007/s11433-020-1617-3
[arXiv:2008.04298 [hep-ph]].

\bibitem{Wang:2014bba}
D.~Wang \textit{et al.} [PVDIS],
Nature \textbf{506}, no.7486, 67-70 (2014)
doi:10.1038/nature12964

\bibitem{Wang:2014guo}
D.~Wang, K.~Pan, R.~Subedi, Z.~Ahmed, K.~Allada, K.~A.~Aniol, D.~S.~Armstrong, J.~Arrington, V.~Bellini and R.~Beminiwattha, \textit{et al.}
Phys. Rev. C \textbf{91}, no.4, 045506 (2015)
doi:10.1103/PhysRevC.91.045506
[arXiv:1411.3200 [nucl-ex]].

\bibitem{Becker:2018ggl}
D.~Becker, R.~Bucoveanu, C.~Grzesik, K.~Imai, R.~Kempf, K.~Imai, M.~Molitor, A.~Tyukin, M.~Zimmermann and D.~Armstrong, \textit{et al.}
doi:10.1140/epja/i2018-12611-6
[arXiv:1802.04759 [nucl-ex]].

\bibitem{Dawson:2018dxp}
S.~Dawson, P.~P.~Giardino and A.~Ismail,
Phys. Rev. D \textbf{99}, no.3, 035044 (2019)
doi:10.1103/PhysRevD.99.035044
[arXiv:1811.12260 [hep-ph]].

\bibitem{Alte:2018xgc}
S.~Alte, M.~K\"onig and W.~Shepherd,
JHEP \textbf{07}, 144 (2019)
doi:10.1007/JHEP07(2019)144
[arXiv:1812.07575 [hep-ph]].

\bibitem{Boughezal:2020uwq}
R.~Boughezal, F.~Petriello and D.~Wiegand,
Phys. Rev. D \textbf{101}, no.11, 116002 (2020)
doi:10.1103/PhysRevD.101.116002
[arXiv:2004.00748 [hep-ph]].

\bibitem{Boughezal:2020klp}
R.~Boughezal, C.~Y.~Chen, F.~Petriello and D.~Wiegand,
[arXiv:2010.06685 [hep-ph]].

\bibitem{Zyla:2020zbs}
P.~A.~Zyla \textit{et al.} [Particle Data Group],
PTEP \textbf{2020}, no.8, 083C01 (2020)
doi:10.1093/ptep/ptaa104

\bibitem{Denner:1991kt}
A.~Denner,
Fortsch. Phys. \textbf{41}, 307-420 (1993)
doi:10.1002/prop.2190410402
[arXiv:0709.1075 [hep-ph]].


\bibitem{Solid1} 
The SOLID collaboration,
"`SoLID (Solenoidal Large Intensity Device) Updated Preliminary Conceptual Design Report''
https://hallaweb.jlab.org/12GeV/SoLID/files/solid-precdr-Nov2019.pdf

\bibitem{Solid2}
D.~Wang, K.~Pan, R.~Subedi, et al.
Nature 506, 67-70 (2014).
https://doi.org/10.1038/nature12964

\bibitem{Solid3}
Addendum for PAC35
Precision Measurement of Parity-violation in Deep Inelastic Scattering Over a Broad Kinematic Range (2009).

  \bibitem{Hobbs:2008mm}
T.~Hobbs and W.~Melnitchouk,
Phys. Rev. D \textbf{77}, 114023 (2008)
doi:10.1103/PhysRevD.77.114023
[arXiv:0801.4791 [hep-ph]].

\bibitem{Mantry:2010ki}
S.~Mantry, M.~J.~Ramsey-Musolf and G.~F.~Sacco,
Phys. Rev. C \textbf{82}, 065205 (2010)
doi:10.1103/PhysRevC.82.065205
[arXiv:1004.3307 [hep-ph]].

\bibitem{Buckley:2012tc}
M.~R.~Buckley and M.~J.~Ramsey-Musolf,
Phys. Lett. B \textbf{712} (2012), 261-265
doi:10.1016/j.physletb.2012.04.077
[arXiv:1203.1102 [hep-ph]].

\bibitem{GonzalezAlonso:2012jb}
M.~Gonz\'alez-Alonso and M.~J.~Ramsey-Musolf,
Phys. Rev. D \textbf{87} (2013) no.5, 055013
doi:10.1103/PhysRevD.87.055013
[arXiv:1211.4581 [hep-ph]].

\bibitem{Aaboud:2017buh}
M.~Aaboud \textit{et al.} [ATLAS],
JHEP \textbf{10} (2017), 182
doi:10.1007/JHEP10(2017)182
[arXiv:1707.02424 [hep-ex]].

\bibitem{Jenkins:2013zja}
E.~E.~Jenkins, A.~V.~Manohar and M.~Trott,
JHEP \textbf{10}, 087 (2013)
doi:10.1007/JHEP10(2013)087
[arXiv:1308.2627 [hep-ph]].

\bibitem{Jenkins:2013wua}
E.~E.~Jenkins, A.~V.~Manohar and M.~Trott,
JHEP \textbf{01}, 035 (2014)
doi:10.1007/JHEP01(2014)035
[arXiv:1310.4838 [hep-ph]].

\bibitem{Alonso:2013hga}
R.~Alonso, E.~E.~Jenkins, A.~V.~Manohar and M.~Trott,
JHEP \textbf{04}, 159 (2014)
doi:10.1007/JHEP04(2014)159
[arXiv:1312.2014 [hep-ph]].

\bibitem{Aad:2016zzw}
G.~Aad \textit{et al.} [ATLAS],
JHEP \textbf{08}, 009 (2016)
doi:10.1007/JHEP08(2016)009
[arXiv:1606.01736 [hep-ex]].


\bibitem{Anthony:2005pm}
P.~L.~Anthony \textit{et al.} [SLAC E158],
Phys. Rev. Lett. \textbf{95} (2005), 081601
doi:10.1103/PhysRevLett.95.081601
[arXiv:hep-ex/0504049 [hep-ex]].

\bibitem{Wood:1997zq}
C.~S.~Wood, S.~C.~Bennett, D.~Cho, B.~P.~Masterson, J.~L.~Roberts, C.~E.~Tanner and C.~E.~Wieman,
Science \textbf{275} (1997), 1759-1763
doi:10.1126/science.275.5307.1759

\bibitem{Androic:2017ppx}
D.~Androic, D.~S.~Armstrong, A.~Asaturyan, T.~Averett, J.~Balewski, K.~Bartlett, J.~Beaufait, R.~S.~Beminiwattha, J.~Benesch and F.~Benmokhtar, \textit{et al.}
EPJ Web Conf. \textbf{137} (2017), 08005
doi:10.1051/epjconf/201713708005

\bibitem{Akhundov:1985fc}
A.~A.~Akhundov, D.~Y.~Bardin and T.~Riemann,
Nucl. Phys. B \textbf{276}, 1-13 (1986)
doi:10.1016/0550-3213(86)90014-3

\bibitem{Beenakker:1988pv}
W.~Beenakker and W.~Hollik,
Z. Phys. C \textbf{40}, 141 (1988)
doi:10.1007/BF01559728

\end{thebibliography}
\end{document}